\documentclass[twocolumn,showpacs,aps,prl,superscriptaddress,preprintnumbers,amsmath,amssymb,floatfix]{revtex4}

\usepackage{graphicx}% Include figure files
\usepackage{dcolumn}% Align table columns on decimal point
\usepackage{bm}% bold math

\usepackage{relsize}
\usepackage{xspace}
\def\babar{\rm{\slshape B\kern-0.1em{\smaller A}\kern-0.1em
    B\kern-0.1em{\smaller A\kern-0.2em R}}}
\def\pep2{PEP-II}
\def\D0bar{\kern 0.2em\overline{\kern -0.2em D}{\kern 0.1em}\xspace^0}

\newcommand{\BABARPubYear}    {04}
\newcommand{\BABARPubNumber}  {025}
\newcommand{\SLACPubNumber} {10654}

\newcommand{\gevc}{\ensuremath{{\mathrm{\,Ge\kern -0.1em V\!/}c}}\xspace}
\newcommand{\mevc}{\ensuremath{{\mathrm{\,Me\kern -0.1em V\!/}c}}\xspace}
\newcommand{\gevcc}{\ensuremath{{\mathrm{\,Ge\kern -0.1em V\!/}c^2}}\xspace}
\newcommand{\mevcc}{\ensuremath{{\mathrm{\,Me\kern -0.1em V\!/}c^2}}\xspace}

\begin{document}

\preprint{BABAR-PUB-\BABARPubYear/\BABARPubNumber}

\preprint{SLAC-PUB-\SLACPubNumber}

\title{\boldmath Search for $D^0$-$\D0bar$ Mixing Using Semileptonic Decay Modes}

\author{B.~Aubert}
\author{R.~Barate}
\author{D.~Boutigny}
\author{F.~Couderc}
\author{J.-M.~Gaillard}
\author{A.~Hicheur}
\author{Y.~Karyotakis}
\author{J.~P.~Lees}
\author{V.~Tisserand}
\author{A.~Zghiche}
\affiliation{Laboratoire de Physique des Particules, F-74941 Annecy-le-Vieux, France }
% ...
\collaboration{The \babar\ Collaboration}
\noaffiliation
\date{\today}

\begin{abstract}
Based on an 87-fb$^{-1}$ dataset collected by the \babar\ detector at the PEP-II asymmetric-energy $B$-Factory, a search for $D^{0}$--$\overline{D}^{0}$ mixing has been made using the semileptonic decay modes $D^{*+} \rightarrow \pi^{+} D^{0}, D^{0} \rightarrow K^{(*)}e\nu$ (+c.c.). The use of these modes allows unambiguous flavor tagging and a combined fit of the $D^{0}$ decay time and $D^{*+}$--$D^{0}$ mass difference ($\Delta M$) distributions. The high-statistics sample of unmixed semileptonic $D^{0}$ decays is used to model the $\Delta M$ distribution and time-dependence of mixed events directly from the data. Neural networks are used to select events and reconstruct the $D^{0}$. A result consistent with no charm mixing has been obtained, $R_{\rm{mix}}=0.0023 \pm 0.0012 \pm 0.0004$. This corresponds to an upper limit of $R_{\rm{mix}}<0.0042$ (90\% CL).
\end{abstract}


\maketitle

In the Standard Model (SM), $D^0$-$\D0bar$ mixing typically proceeds through box diagram loops involving down-type quarks. This results in very effective GIM and CKM suppressions of the mixing rate. The expected rate, relative to the unmixed rate, of $D^0$-$\D0bar$ mixing through SM box \cite{Datta:1985jx} and di-penguin \cite{Petrov:1997ch} diagrams is $\mathcal{O}(10^{-8}-10^{-10})$, many orders of magnitude below the current experimental sensitivity of $\mathcal{O}(10^{-3}$) \cite{Burdman:2003rs}. Possible enhancements to the SM mixing rate involve non-perturbative effects, and predictions range over many orders of magnitude \cite{Petrov:2003un} approximately bounded by the box diagram rate and the experimental sensitivity. New physics charm mixing predictions span the same large range \cite{Petrov:2003un} and, therefore, the presence of a mixing signal alone would not be a clear indication of new physics. However, the current experimental bounds on charm mixing already constrain new physics models.

Charm mixing is generally characterized by two dimensionless parameters, $x\equiv\Delta m/\Gamma$ and $y\equiv\Delta\Gamma/2\Gamma$, where $\Delta m=m_{2}-m_{1}$ ($\Delta\Gamma=\Gamma_{2}-\Gamma_{1}$) is the mass (width) difference between the two neutral $D$ mass eigenstates and $\Gamma$ is the average width. If either $x$ or $y$ is non-zero, then $D^0$-$\D0bar$ mixing will occur.

The time evolution of a neutral $D$ meson depends on the type of final state into which it decays, since doubly Cabibbo-suppressed (DCS) decays to the same final states as mixed events allow for interference between the mixing and DCS amplitudes \cite{Blaylock:1995ay}. Semileptonic decays, for which there are no DCS amplitudes, have a particularly simple time-dependence,
\begin{equation}
T_{\rm{mix}}(t)\cong T_{\rm{unmix}}(t) \ \frac{x^{2}+y^{2}}{4} \ \left(\frac{t}{\tau_{D^{0}}}\right)^{2},
\label{eqn:semileptime}
\end{equation}
\noindent where $t$ is the proper time of the $D^0$ decay, $T_{\rm{unmix}}(t)\propto e^{-t/\tau_{D^{0}}}$, and the approximation is valid in the limit of small mixing rates. Sensitivity to $x$ and $y$ individually is lost with semileptonic final states. The time-integrated mixing rate $R_{\rm{mix}}$ relative to the unmixed rate is
\begin{equation}
R_{\rm{mix}}= \frac{x^{2}+y^{2}}{2}.
\label{eqn:mixrate}
\end{equation}

We present a measurement of $R_{\rm{mix}}$ using an 87-fb$^{-1}$ data sample collected on and just below the $\Upsilon(4S)$ resonance with the \babar\ detector \cite{Aubert:2001tu} at the PEP-II asymmetric-energy $e^{+}e^{-}$ storage ring. A full \babar\ detector Monte Carlo simulation based on GEANT4 \cite{Agostinelli:2002hh} is used to study background sources and to develop methods to reconstruct neutral $D$ mesons decaying to semileptonic final states.

Charged tracks are measured in a five-layer silicon vertex tracker (SVT) surrounded by a cylindrical wire drift chamber (DCH). Charged hadrons are identified through ${\rm d}E/{\rm d}x$ measurements in the tracking volume and in a ring-imaging Cherenkov detector (DIRC) surrounding the DCH. Identified charged $K$ purities averaged across all momenta are $\sim$99\%, with $\sim$80\% reconstruction efficiency. Relative charge asymmetries between $K^{+}$ and $K^{-}$ in both misidentification rate and reconstruction efficiency are at most a few percent across all momenta and contribute negligibly to the total error on the mixing rate. Charge conjugation is implied unless explicitly stated otherwise.

Electromagnetic showers from electrons and photons are detected by CsI crystals arrayed in an electromagnetic calorimeter (EMC) located between the DIRC and the superconducting solenoid. Electrons are differentiated from other charged tracks by the intrinsic differences between hadronic, muonic and electronic energy depositions in the EMC. As with charged $K$ candidates, there are relative charge asymmetries of at most a few percent in both misidentification rate and reconstruction efficiency between positrons and electrons, which have a negligible contribution to the total error on the mixing rate. Electrons and positrons arising from gamma conversions are eliminated from the pool of tracks used to reconstruct $D^{0}$ candidates.

Neutral $D$ candidates are selected by reconstructing the decay chain $D^{*+} \rightarrow \pi^{+} D^{0}, D^{0} \rightarrow K^{(*)}e\nu$. There are no essential differences for this analysis between the $K$ and $K^{*}$ modes, either theoretically or empirically, and thus no attempt is made to reconstruct the $K^{*}$ --- its charged $K$ daughter is treated as if it was a direct daughter of the $D^{0}$. The charge of the pion daughter of the charged $D^{*}$ identifies the production flavor of the neutral $D$, while the charge of the electron identifies the decay flavor. These charges are the same for unmixed decays and different for mixed decays, denoted as \emph{right-sign} (RS) and \emph{wrong-sign} (WS) decays, respectively.

The measured mixing rate is parameterized as $R_{\rm{mix}}=n_{\rm{WS}}/n_{\rm{RS}}$. We fit the number of RS (WS) signal decays, $n_{\rm{RS}}$ ($n_{\rm{WS}}$), with a likelihood combining the $D^{*+}$-$D^{0}$ mass difference ($\Delta M$) distribution with the unmixed and mixed (Eq.~\ref{eqn:semileptime}) decay time distributions, respectively. The high-statistics sample of RS signal events is used to model the $\Delta M$ distribution and the time-dependence of mixed events directly from the data. To avoid potential bias, we perform a blind analysis in which the event selection criteria, and procedures for fitting the data and extracting an upper limit (UL), are determined prior to examining the WS signal region $\Delta M$ and decay time distributions in the data.

Identified $K$ and $e$ candidates of opposite charges are combined to create neutral $D$ candidate decay vertices. Only candidates with vertex fit probability $>$ 0.01 and mass $<$ 1.82~GeV/$c^{2}$ are retained. The value of the mass cut is chosen to exclude all hadronic two-body $D^{0}$ decays. The average PEP-II interaction point (IP), measured on a run-to-run basis using Bhabha and $\mu^{+} \mu^{-}$ events, is taken as the production point of $D^{0}$ candidates. The transverse $D^{0}$ decay time is used due to the relative narrowness of, and small errors on, the IP in the transverse ($r$-$\phi$) plane, as opposed to the IP's longitudinal ($z$) profile. Because of the very small transverse distance traveled by $B$ mesons prior to decay, there are no significant differences between the decays of charmed parents arising from the continuum (\emph{prompt} events) and those coming from the decay of $B$ mesons, other than the harder $D^{0}$ momentum spectrum from prompt events.

The pions from $D^{*+}$ decays are relatively soft tracks with $p^{*}_{\pi}<450$~MeV/$c$, where the asterisk denotes a parameter measured in the $\Upsilon(4S)$ center-of-mass frame. Charged tracks identified as either a charged $K$ or $e$ candidate are not considered as $\pi$ candidates. To reject poorly reconstructed tracks, $\pi$ candidates are required to have six or more SVT hits, with $\geq2$ $r$-$\phi$ layer hits, $\geq2$ $z$ layer hits, at least one hit on the inner three $r$-$\phi$ layers and at least one hit on the inner three $z$ layers. Pion candidates are also refit with the IP as a constraint and accepted only if the refit probability is greater than 0.01.

The individual $K$, $e$ and $\pi$ tracks, the $K$-$e$ vertex, and the event thrust are used to reconstruct the three orthogonal components of the $D^{0}$ momentum vector with three JetNet 3.4 \cite{Peterson:1994nk} neural networks (NN's). Each NN has two hidden layers, and is trained and tested with a large sample $(\mathcal{O}(10^{5}))$ of simulated signal events. The following vector inputs to the NN's are used: the momentum of the $K$-$e$ pair constrained by a vertex fit [${\bf p}^{*}$($K$-$e$)], ${\bf p}^{*}$($\pi$) and the event thrust vector ${\bf T}$ (calculated using all charged and neutral candidates except the $K$ and $e$ candidates). In addition, four opening angles among the above parameters are used: $\theta^{*}({\bf p}_{K\rm{-}e},{\bf T})$, $\theta^{*}({\bf p}_{K},{\bf p}_{e})$, $\theta^{*}({\bf p}_{\pi},{\bf T})$, and $\theta^{*}({\bf p}_{K\rm{-}e},{\bf p}_{\pi})$. The vector inputs are separated into their respective orthogonal components (azimuthal/polar angles and magnitude) and, additionally using the four scalar inputs, separate NN's for $p^{*}_{\phi}(D^{0})$, $p^{*}_{\theta}(D^{0})$, and $|{\bf p}^{*}(D^{0})|$ are trained with the seven input parameters. The residuals between the true ${\bf p}^{*}$($D^{0}$) direction for simulated signal events and the NN output are unbiased and Gaussian-distributed with $\sigma \sim$110~mrad. The distribution of momentum magnitude residuals shows an error of $\sigma_{p}/p\sim$10\% for a typical signal event.

The $D^{*+}$--$D^{0}$ mass difference is calculated using the $D^{0}$ candidate and the tagging pion candidate. The signal $\Delta M$ distribution peaks $\sim$3~MeV/$c^{2}$ below the PDG \cite{shortPDBook} value of 145.4~MeV/$c^{2}$ and has a RMS width of $\sim$2.2~MeV/$c^{2}$. The shift in the peak introduces no bias into the measurement of the mixing rate and is due to the use of tagging pion information in reconstructing the $D^{0}$ momentum. A $\sim$80~MeV/$c^{2}$ wide $\Delta M$ sideband is retained from 160~MeV/$c^{2}$ to 240~MeV/$c^{2}$ in order to characterize the level of background events.

The transverse momentum of a $D^{0}$ candidate, and the projections of the IP and $K$-$e$ vertex loci on the $r$-$\phi$ plane, are used to calculate the candidate's proper decay time. The error on the decay time, calculated using only the errors on the IP and $K$-$e$ vertex, is typically 0.6$\tau_{D^{0}}$, where $\tau_{D^{0}}$ is the nominal mean $D^{0}$ lifetime. The contribution of the ${\bf p}^{*}$($D^{0}$) estimator to the total decay time uncertainty is negligible. Poorly reconstructed events with calculated decay time uncertainties greater than 2$\tau_{D^{0}}$ are discarded. Only events with decay times between $-12\tau_{D^{0}}$ and $16\tau_{D^{0}}$ are retained.

In addition to the above criteria, events are selected using a neural network trained to distinguish prompt charm signal from background events. The event selector NN uses a five-element input vector: $p^{*}_{K\rm{-}e}$, $p^{*}_{\pi}$, $|{\bf T}|$, $\theta^{*}({\bf p}_{K\rm{-}e},{\bf T})$, and $\theta^{*}({\bf p}_{K},{\bf p}_{e})$. It has a single hidden layer of nine nodes and is also constructed using JetNet~3.4. A cut is made on the event selector NN output such that the statistical sensitivity to mixed charm events is optimized for the 87-fb$^{-1}$ dataset used.

Events with multiple candidates meeting all of the above selection criteria but which share tracks are discarded, leading to a loss of $\sim$10\% of signal events passing all other cuts. There are no differences in RS and WS reconstruction efficiencies in the final event selection.

Backgrounds predominantly come from prompt charm events, with minor contributions from $uds$ and $b\overline{b}$ events. The backgrounds from misidentified charged particle species are negligible. Nearly all background events come from $D^{0}$ and $D^{+}$ semileptonic decays to final states including both a charged $K$ and an $e$ that are combined with a random $\pi^{+}$, and truly random combinatorics in which the $K$ and $e$ do not share a common charm parent. The decay time distributions of background charm parents differ somewhat from those of well-reconstructed charm parents because the ${\bf p}^{*}$($D^{0}$) estimator returns a correct momentum estimate only for signal events. The decay time probability density functions (PDF's) for these charm backgrounds are thus not fit from the data but, instead, are derived from simulated events. The decay time distribution for truly random (i.e., zero-lifetime) events is fit in the data using a sum of three Gaussians.

The $\Delta M$ backgrounds qualitatively fall into two categories --- those which peak under the signal $\Delta M$ distribution and those which do not. Both RS and WS events have small peaking $\Delta M$ backgrounds. In the WS sample, the peaking component arises from semi-electronic $D^{+}$ decays with a $K^{*}$ daughter --- these events are fit using the $D^{+}$ decay time PDF and are thus differentiated from mixed signal events. In the RS sample, the fraction of background events sharing the RS unmixed signal PDF is estimated from fits to simulated datasets and a correction of $\sim$4\% is made to the fit number of RS signal decays to obtain the actual number. The RS $D^{+}$ $\Delta M$ background peaks well away from the $\Delta M$ signal region and its shape is fixed from simulated events. The shape is floated as a systematic check and there is a negligible effect on the final value of $R_{\rm{mix}}$.

Non-peaking combinatoric $\Delta M$ backgrounds are modeled from the data by combining $K$-$e$ vertex and $\pi$ candidates from different events (``event mixing"). Events are selected from the off-resonance data such that a $K$-$e$ vertex drawn from one event may be combined with a $\pi$ candidate from any other event. Approximately 70,000 off-resonance data events are used to generate $\mathcal{O}(10^{6})$ RS and WS pseudo-events passing the final event selection criteria. A Kolmogorov-Smirnov test \cite{kolmogorov} shows that the event-mixed RS and WS $\Delta M$ distributions have a $\sim$92\% probability of sharing the same underlying parent distribution. The shape of the PDF associated with the non-peaking combinatoric $\Delta M$ background is taken from the combined event-mixed RS and WS $\Delta M$ distributions and denoted by $H(\Delta M)$.

An initial fit to the RS data is used to extract the unmixed signal mean lifetime, shape of the signal $\Delta M$ distribution, and number of RS signal events. The mixed WS signal PDF parameters are taken from this high-statistics fit. The full extended likelihood function is:
\begin{equation}
\mathcal{L} = \frac{\lambda^{N}e^{-\lambda}}{N!}\prod_{i=1}^N\sum_{j=1}^m f_{j}\mathcal{P}_{j},
\end{equation}
\noindent where $f_{j}$ ($n_{j}$) is the fraction (number) of events assigned to event class $j$, $\lambda=\sum n_{j}$ is the fit total number of candidates, $\mathcal{P}_{j}$ is the value of the PDF for event class $j$, and the extended product runs over all N events. The PDF's for RS and WS signal and background classes are detailed below. The shapes of all PDF's are fit in the data unless otherwise noted. All fits are done as unbinned maximum likelihood fits.

Four classes of events are fit in the RS data: unmixed signal events ($n_{\rm{unmix}}$), background $D^{0}$ decays ($n_{\rm{bkgd D^{0}}}$), RS $D^{+}$ decays ($n_{\rm{RS D^{+}}}$) and completely random three-track combinations ($n_{\rm{zero-life}}$). The RS signal $\Delta M$ shape is fit to a threshold function with a power-law turn-on and an exponentially decaying tail, while the decay time distribution is fit as an exponential decay with floating decay constant convoluted with a resolution model ($G$) taken from simulated signal events. The RS signal $\Delta M$ and decay time combined PDF is:
\begin{equation}
\mathcal{P}_{\rm{unmix}}=u^{\gamma}\exp(\sum^{3}_{i=1}\alpha_{i}u^{i}) \cdot [e^{-t/\tau_{D^{0}}}\theta(ct>0) \otimes G],
\label{eqn:pdfrss}
\end{equation}
\noindent where $u=\Delta M - M_{\pi^{+}}$, and $M_{\pi^{+}}$ is fixed to the nominal PDG value \cite{shortPDBook}. The four parameters of the signal $\Delta M$ PDF ($\gamma$, $\alpha_{i}$), $\tau_{D^{0}}$ and the unmixed signal population are allowed to float in the RS fit.

The background $D^{0}$ event class uses the combined PDF:
\begin{equation}
\mathcal{P}_{\rm{bkgd D^{0}}}=H(\Delta M) \cdot T_{D^{0}}(t),
\label{eqn:pdfd0}
\end{equation}
\noindent where $T_{D^{0}}(t)$ represents the shape of the decay time distribution for this event class derived from simulated events. The RS $D^{+}$ event class has the combined PDF:
\begin{equation}
\mathcal{P}_{\rm{RS D^{+}}}=u^{\zeta}\exp(\xi_{1}u+\xi_{2}u^{2}+\xi_{3}u^{3}) \cdot T_{D^{+}}(t),
\label{eqn:pdfdplus}
\end{equation}
\noindent where $\zeta$ and $\xi_{i}$ are fixed from fits to simulated events, and $T_{D^{+}}(t)$ represents the shape of the decay time distribution for this event class derived from simulated events. Finally, zero-lifetime events use the combined PDF:
\begin{equation}
\mathcal{P}_{\rm{zero-life}}=H(\Delta M) \cdot \sum_{i=1}^3f_{i}\exp\left[-\frac{1}{2}\left(\frac{\mu-ct}{\sigma_{i}}\right)^{2}\right],
\label{eqn:pdfzerolife}
\end{equation}
\noindent with the constraint $f_{3}=1-f_{1}-f_{2}$. Fifteen parameters are floated in the RS fit: the signal $\Delta M$ shape parameters and mean $D^{0}$ lifetime; the mean, widths and relative fractions of the zero-lifetime triple Gaussian; and the population of the four RS event classes.

\begin{figure}
\includegraphics[width=\linewidth]{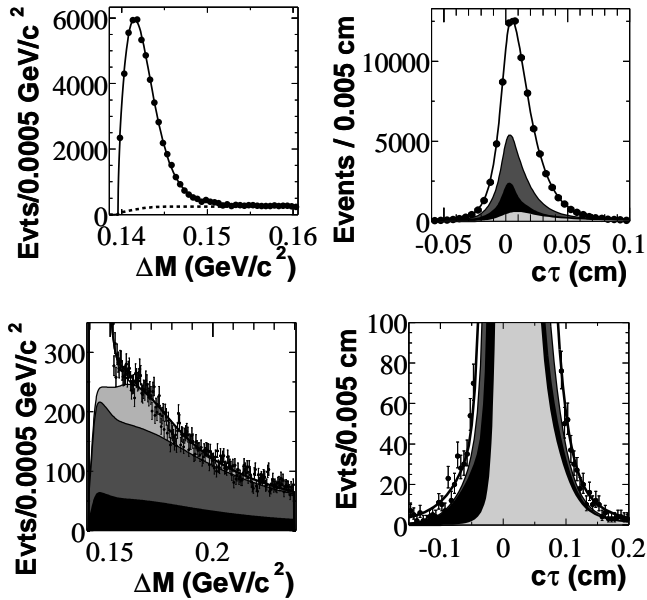}
\vspace{-8 mm}\caption{\label{fig:rsfit}$\Delta M$ (left) and decay time (right) projections of fit (solid lines) to RS data (points): (top left) $\Delta M$ signal region --- unmixed signal (above dashed line), background (dashed line); (bottom left) magnified vertical scale $\Delta M$ full fit region --- unmixed signal (white), $D^{+}$ background (light grey), $D^{0}$ background (dark grey), zero-lifetime background (black); (top right) decay time signal region --- signal and background components as in bottom left plot; (bottom right) magnified vertical scale decay time full fit region --- signal and background components as in bottom left plot of this figure.}
\end{figure}

Figure~\ref{fig:rsfit} shows the projections of the RS fit model onto the RS data. After correcting for peaking RS background events as described above, we find $n_{\rm{unmix}}=49620\pm265$~events. The fit value for the unmixed $D^{0}$ mean lifetime is, within its statistical error of $\sim$0.6\%, consistent with the current PDG \cite{shortPDBook} value.

\begin{figure}
\includegraphics[width=\linewidth]{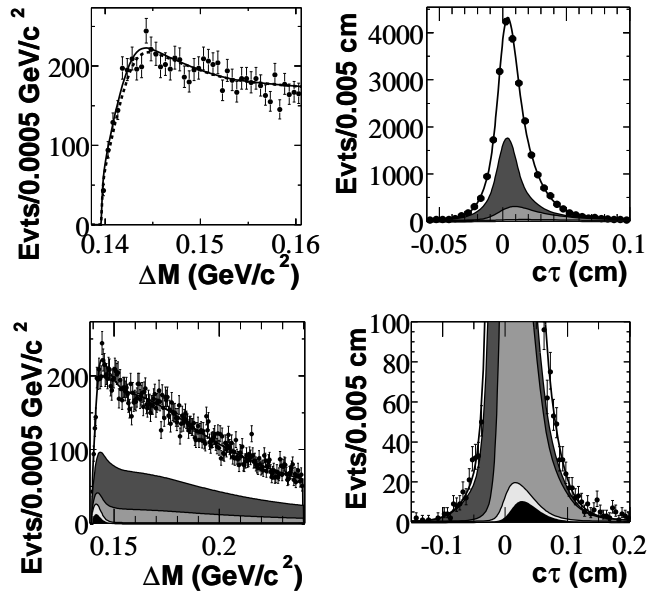}
\vspace{-8 mm}\caption{\label{fig:wsfit}$\Delta M$ (left) and decay time (right) projections of fit (solid lines) to WS data (points): (top left) $\Delta M$ signal region --- mixed signal (above dashed line), background (dashed line); (bottom left) $\Delta M$ full fit region --- $D^{0}$ background (white), zero-lifetime background (dark grey), non-peaking $D^{+}$ background (intermediate grey), peaking $D^{+}$ background (light grey), mixed signal (black); (top right) decay time signal region --- signal and background components as in bottom left plot; (bottom right) magnified vertical scale decay time full fit region --- signal and background components as in bottom left plot.}
\end{figure}

Five classes of events are fit in the WS data: mixed signal events ($n_{\rm{mix}}$), background $D^{0}$ decays ($n_{\rm{bkgdD^{0}}}$), peaking WS $D^{+}$ decays ($n_{\rm{pkng D^{+}}}$), non-peaking WS $D^{+}$ decays ($n_{\rm{non-pkng D^{+}}}$) and completely random three-track combinations ($n_{\rm{zero-life}}$). WS signal events share the sharply peaked $\Delta M$ distribution of the RS signal, but have the modified lifetime distribution given by Eq.~\ref{eqn:semileptime}. The full form of the WS mixed signal PDF is:
\begin{equation}
\mathcal{P}_{\rm{mix}}=u^{\gamma}\exp(\sum^{3}_{i=1}\alpha_{i}u^{i}) \cdot [e^{-t/\tau_{D^{0}}}\theta(ct>0) \otimes G],
\label{eqn:pdfwss}
\end{equation}
\noindent where $\alpha_{i}$, $\gamma$ and $\tau_{D^{0}}$ are fixed to the values found in the RS unmixed fit. The background $D^{0}$ and random combinatoric event classes use the combined likelihoods shown in Eqs.~\ref{eqn:pdfd0} and~\ref{eqn:pdfzerolife}, respectively. The peaking WS $D^{+}$ event class uses the fixed signal $\Delta M$ PDF and the $D^{+}$ lifetime model:
\begin{equation}
\mathcal{P}_{\rm{pkng D^{+}}}=u^{\gamma}\exp(\alpha_{1}u+\alpha_{2}u^{2}+\alpha_{3}u^{3}) \cdot T_{D^{+}}(t).
\label{eqn:pdfpkngdplus}
\end{equation}
Non-peaking WS $D^{+}$ events use the likelihood:
\begin{equation}
\mathcal{P}_{\rm{non-pkng D^{+}}}=H(\Delta M) \cdot T_{D^{+}}(t).
\label{eqn:pdfnonpkngdplus}
\end{equation}
Eleven parameters are floated in the WS fit: the mean, widths and relative fractions of the zero-lifetime triple Gaussian and the populations of the five WS event classes. Figure~\ref{fig:wsfit} shows the projections of the WS fit model onto the WS data. We find $n_{\rm{mix}}=114\pm61$~events, leading to a value of $R_{\rm{mix}}=0.0023\pm0.0012$~(stat). Fits to toy Monte Carlo data sets show that, for an assumed zero mixing rate, a fit number of mixed events greater than the result here is likely to occur in about 5\% of experiments. No significant asymmetries are seen when the WS dataset is divided and fit based on the production flavor of the $D^0$-$\D0bar$.

Goodness-of-fit is checked by comparing the minimized negative log likelihood (NLL) values of the RS and WS data fits with NLL distributions generated from toy MC --- the data-fit NLL values lie well within the range predicted by the toy fits. The WS fit model is also tested for bias and correct error scaling with fits to simulated datasets containing 0, 50 and 100 mixed WS events ($R_{\rm{mix}}\sim 0$, 0.001, 0.002, respectively), and no evidence of bias or improperly scaled errors is seen.

The systematic error includes variations of the WS combined mixed signal PDF, random combinatoric $\Delta M$ PDF shape, signal decay time resolution model, and background $D^{0}$ and $D^{+}$ decay time PDF's. By far, the dominant systematic is the statistical precision with which the RS $\Delta M$ PDF is known. This systematic is characterized by sequentially varying by $\pm1\sigma$ one of the signal $\Delta M$ shape parameters floated in the RS fit, along with the mean $D^{0}$ lifetime, and refitting the WS data. The resulting deviations from the central value of $n_{\rm{mix}}$ obtained in the nominal fit to the WS data are summed in quadrature, taking into account correlations among the fit parameters. Table~\ref{tab:sys} shows the error on $R_{\rm{mix}}$ attributable to each source of systematic error listed above. There are no significant effects on $R_{\rm{mix}}$ attributable to the choice of vertexing algorithm, IP, $K$ or $e$ particle identification, NN event selector cut, decay time error cut, signal resolution model or $\Delta M$ sideband cut. Taking the total systematic error as the sum in quadrature of the values in Table~\ref{tab:sys}, $\sigma^{sys}_{R_{\rm{mix}}}=0.0004=0.34\sigma^{stat}_{R_{\rm{mix}}}$.

\begin{table}
\caption{\label{tab:sys}Systematic variations in the value of $R_{\rm{mix}}$.}
\begin{tabular}{|l|c|} \hline
Varied parameter     & \emph{$|\Delta R_{\rm{mix}}|$} \\ \hline
 Mixed $\Delta M$ PDF                  & 0.00032 \\
 Mixed decay time PDF                  & 0.00008 \\
 Random combinatoric $\Delta M$ shape  & 0.00015 \\
 Bkgd $D^{0}$ decay model              & 0.00016 \\
 Bkgd $D^{+}$ decay model              & 0.00012 \\ \hline
\end{tabular}
\end{table}

A scan of the change in NLL for $n_{\rm{mix}}$ values in the region surrounding the fit minimum is used to calculate upper limits. By construction, the NLL scan includes only the statistical error of the fit --- the systematic error is included as a small perturbation on the values of $\Delta$NLL used to establish confidence intervals. The total error is taken as the sum in quadrature of the statistical and systematic errors, $\sigma^{total} = \sqrt{1+0.34^{2}} \hspace{1.5 mm} \sigma^{stat}=1.06 \hspace{1.5 mm} \sigma^{stat}$. The 95\% CL UL is taken as the value of $n_{\rm{mix}}$ where the NLL value changes from its minimum by $\Delta\rm{NLL}=(0.5)(1.06)(1.96^{2})(0.97)=1.97$, where a one-sigma change is $\Delta\rm{NLL}=0.5$, which yields $R_{\rm{mix}}$$<$0.0046 (95\% CL). The factor of ``0.97" in the preceding expression for the UL arises from the fraction of the \emph{a posteriori} distribution of $n_{\rm{mix}}$ lying in the physical region. A similar calculation shows $R_{\rm{mix}}$$<$0.0042 at the 90\% CL. The relatively small error ($\sim$0.5\%) on $n_{\rm{RS}}$ is negligible and has been ignored.

In summary, we have obtained a value of $R_{\rm{mix}}$ consistent with no mixing and set a limit of $R_{\rm{mix}}$$<$0.0042 (0.0046) at the 90\% (95\%) CL. This is slightly lower than the previous best published limit using semileptonic decays of $R_{\rm{mix}}<0.005$ (90\% CL) set by E791 \cite{Aitala:1996vz}. However, the positive result obtained for $n_{\rm{mix}}$ obscures the actual sensitivity of the analysis. The current best published UL on $R_{\rm{mix}}$ ($<0.0016$, 95\% CL) comes from the \babar\ hadronic $K\pi$ mixing analysis \cite{Aubert:2003ae}. Had the central value of $n_{\rm{mix}}$ been observed here to be zero, the 90\% CL (95\%) UL for the mixing rate would have been 0.0016 (0.0020), comparable to both the hadronic \babar\ result and a recent semileptonic FOCUS analysis \cite{Hosack:2002hr}.

\begin{acknowledgments}
We are grateful for the excellent luminosity and machine conditions
provided by our \pep2\ colleagues,
and for the substantial dedicated effort from
the computing organizations that support \babar.
The collaborating institutions wish to thank
SLAC for its support and kind hospitality.
This work is supported by
DOE
and NSF (USA),
NSERC (Canada),
IHEP (China),
CEA and
CNRS-IN2P3
(France),
BMBF and DFG
(Germany),
INFN (Italy),
FOM (The Netherlands),
NFR (Norway),
MIST (Russia), and
PPARC (United Kingdom).
Individuals have received support from CONACyT (Mexico), A.~P.~Sloan Foundation,
Research Corporation,
and Alexander von Humboldt Foundation.
\end{acknowledgments}

\bibliography{dmix}

\end{document}